\documentclass[aps, pre, twocolumn, superscriptaddress]{revtex4}
\usepackage{bm}
\usepackage{amsmath}
\usepackage{amsfonts}
\usepackage{amsthm}
\usepackage{amssymb}
\usepackage[pdftex]{graphicx}
\usepackage[utf8]{inputenc}

\begin{document}

\newcommand{\defeq}{\mathrel{\mathop:}=}

\title{Computational Statistical Mechanics of a confined, three-dimensional Coulomb gas}

\author{Sergio Davis}
\email{sergio.davis@cchen.cl}

\affiliation{Comisión Chilena de Energía Nuclear, Casilla 188-D, Santiago, Chile}
\affiliation{Departamento de F\'isica, Facultad de Ciencias Exactas, Universidad Andres Bello. Sazi\'e 2212, piso 7, 8370136, Santiago, Chile.}

\author{Jalaj Jain}
\affiliation{Comisión Chilena de Energía Nuclear, Casilla 188-D, Santiago, Chile}

\author{Biswajit Bora}
\affiliation{Comisión Chilena de Energía Nuclear, Casilla 188-D, Santiago, Chile}
\affiliation{Departamento de F\'isica, Facultad de Ciencias Exactas, Universidad Andres Bello. Sazi\'e 2212, piso 7, 8370136, Santiago, Chile.}

\date{\today}

\begin{abstract}
The thermodynamic properties of systems with long-range interactions is still an ongoing challenge, both from the point of view of theory as well as computer simulation. In this work
we study a model system, a Coulomb gas confined inside a sphere, by using the Wang-Landau algorithm. We have computed the configurational density of states (CDOS), the thermodynamic
entropy and the caloric curve, and compared with microcanonical Metropolis simulations, while showing how concepts such as the configurational inverse temperature can be used to
understand some aspects of thermodynamic behavior. A dynamical multistability behavior is seen at low energies in microcanonical Monte Carlo simulations, suggesting that flat-histogram methods are in fact superior alternatives to traditional simulation in complex systems. \end{abstract}

\maketitle

\section{Introduction}

The thermodynamics of charged particles has remained a challenging subject, mainly due to the long-range nature of the interaction which breaks some assumptions of classical statistical
mechanics. While exact results are common for the two-dimensional Coulomb potential~\cite{Bedanov1994, Samaj2003}, the situation in three dimensions for the unscreened Coulomb potential
is not as clear, even using computer simulation.

One interesting phenomenon in long-range interacting systems is the origin of non-Maxwellian velocity distributions, not only in plasmas but in model systems such as the Hamiltonian
mean-field (HMF) model~\cite{Latora2002,Atenas2017}. In such long-range systems different mechanisms have been proposed to explain these distributions, such as Tsallis' nonextensive
statistical mechanics~\cite{Du2004,Tsallis2007}, superstatistics~\cite{Beck2004}, among others. For an isolated system, the velocity distribution of its components is governed by
properties of the interaction potential, more precisely by its configurational density of states (CDOS), as for instance shown by J. R. Ray~\cite{Ray1991b} in small systems. Therefore,
it makes sense to gain some understanding of the behavior of this CDOS for long-range potentials.

In the field of condensed matter physics, on the other hand, the computation of the CDOS for systems with complex interactions (such as proteins) using Monte Carlo methods in
generalized ensembles has recently emerged as a promising alternative~\cite{Rathore2003,Rathore2003a} to molecular dynamics simulations. Nevertheless, we are not aware of a calculation
of the CDOS for pure Coulomb systems in the literature.

In this work, we focus on the thermodynamics of a model system, where charged particles interacting via the unscreened Coulomb potential are confined inside a spherical region. We
present a computation of the CDOS using the Wang-Landau algorithm~\cite{Wang2001}, and from this we determine its equilibrium thermodynamic properties in the canonical and
microcanonical ensembles.

This article is organized as follows. Section \ref{sect_model} defines the interaction energy and the choice of natural units. Sections \ref{sect_microthermo} and \ref{sect_comput}
review the microcanonical formalism in terms of the configurational degrees of freedom, and the implementation of Monte Carlo methods, while Section \ref{sect_results} presents
the main results. Finally we close with some concluding remarks in Section ~\ref{sect_conclud}.

\section{Description of the model}
\label{sect_model}

We will consider a group of $N = N_+ + N_-$ charged particles in three dimensions. The $N$ particles are divided exactly into two equal groups of $N_+=N_-=N/2$ with charge $q_+=e$
and $q_-=-e$, respectively, in order to have exact neutrality. In this system, the Hamiltonian is
\begin{equation}
H(\bm{r}_1, \ldots, \bm{r}_N, \bm{p}_1, \ldots, \bm{p}_N) = \sum_{i=1}^N \frac{\bm{p}_i^2}{2m_i} + \Phi(\bm{r}_1, \ldots, \bm{r}_N),
\label{eq_hamil}
\end{equation}
where $\Phi$ is the electrostatic potential energy, given by
\begin{equation}
\Phi(\bm{r}_1, \ldots, \bm{r}_N) = \frac{1}{2}\sum_{i=1}^N \sum_{j \neq i} \frac{q_i q_j}{4\pi \epsilon_0 |\bm{r}_j-\bm{r}_i|},
\end{equation}
which can be expressed in natural units by defining a natural length unit $r_0$. We can write
\begin{equation}
\Phi(\bm{r}_1, \ldots, \bm{r}_N) = \frac{\phi_0}{2}\sum_{i=1}^N \sum_{j \neq i} \frac{\sigma_i \sigma_j}{|\bm{r}_j-\bm{r}_i|},
\label{eq_phi_spin}
\end{equation}
with $\sigma_i=\pm 1$ and $r_{ij}$ in units of $r_0$, the model now resembling a long-range Ising-type interaction but with mobile ``spins''.

\noindent
The unit of energy corresponds to
\begin{equation}
\phi_0 \defeq \frac{e^2}{4\pi \epsilon_0 r_0}.
\end{equation}

At this point, we will introduce two modifications to the model. First, in order to ``soften'' the interaction at very short distances, we have corrected the
interparticle distance as $$|\bm{r}_j-\bm{r}_i| \rightarrow \max(|\bm{r}_j-\bm{r}_i|, r_0),$$ which avoids the singular behavior at $r_{ij}$=0, making the potential energy bounded.
As shown originally by Fisher and Ruelle~\cite{Fisher1966}, this truncation of the Coulomb potential is one mechanism able to restore stability. Second, because our aim is to describe
an isolated, finite-size system, the particles are confined inside a sphere of radius $R$, so that $|\bm{r}_i| < R$.

\begin{table}[t!]
\begin{center}
\begin{tabular}{|c|c|c|}
\hline
Description & Symbol & Reference value \\
\hline
Length unit & $r_0$ & 0.529 \AA \\
\hline
Energy unit & $\phi_0=e^2/(4\pi\epsilon_0 r_0)$ & 27.211 eV \\
\hline
Temperature unit & $T_0=\phi_0/k_B$ & 315774 K \\
\hline
Number of particles & $N$ & 210 \\
\hline
Confining radius & $R$ & 146 \AA \\
\hline
Particle density & $n=N/(\frac{4}{3}\pi R^3)$ & 8.05$\times$10$^{24}$ m$^{-3}$ \\
\hline
Debye length at $T=T_0$ & $\lambda_D$ & 137.1 \AA \\
\hline
\end{tabular}
\end{center}
\caption{Parameters in physical units for the choice of $r_0$ equal to the Bohr radius.}
\label{tbl_parameters}
\end{table}

In the following we use $N$=210 particles and a confining radius $R=$276 $r_0$. For reference, the values of some parameters of interest in plasma physics are given in
Table~\ref{tbl_parameters}, for the choice of $r_0$ equal to the Bohr radius, $r_0$=0.529 \AA. In this case, the system is denser than magnetic confinement plasmas but less dense
than inertial fusion plasmas~\cite{Bellan2006}.

\section{Microcanonical thermodynamics}
\label{sect_microthermo}

For a system with Hamiltonian given by Eq. \ref{eq_hamil}, because the form of the kinetic energy is universal, the CDOS becomes the key quantity for the thermodynamics in steady
states. We will consider a steady state described by the ensemble function $\rho(E)$, such that
\begin{equation}
P(\bm{R}, \bm{P}|\mathcal{S})=\rho(H(\bm{R}, \bm{P})),
\end{equation}
where $\bm{R}=(\bm{r}_1, \ldots, \bm{r}_N)$ and $\bm{P}=(\bm{p}_1, \ldots, \bm{p}_N)$.

In such an ensemble, the expectation of any function $g(H)$ of the energy can be computed as
\begin{align}
\Big<g\Big>_{\mathcal{S}} & = \int d\bm{R}d\bm{P}\;g(H(\bm{R}, \bm{P})) \;P(\bm{R}, \bm{P}|\mathcal{S}) \nonumber \\
                          & = \int d\bm{R}d\bm{P}\; g(K(\bm{P})+\Phi(\bm{R}))\rho(K(\bm{P})+\Phi(\bm{R})) \nonumber \\
                          & = \int d\phi\; dK \;\Omega_K(K) \mathcal{D}(\phi)\cdot g(K+\phi)\rho(K+\phi),
\end{align}
where we have introduced $\Omega_K(K)$, the density of states of the classical ideal gas,
\begin{align}
\Omega_K(K) & \defeq \int d\bm{p}_1\ldots d\bm{p}_N\;\delta\left(\sum_{i=1}^N\frac{\bm{p}_i^2}{2m_i} -K\right) \nonumber \\
            & = \Omega_0\;K^{\frac{3N}{2}-1},
\label{eq_omega_k}
\end{align}
and $\mathcal{D}(\phi)$, the configurational density of states, defined in our case by the multidimensional integral
\begin{equation}
\mathcal{D}(\phi) \defeq \int_{\mathcal{S}(R)}\hspace{-5pt} d\bm{r}_1\ldots d\bm{r}_N \;\delta(\phi - \Phi(\bm{r}_1, \ldots, \bm{r}_N)),
\end{equation}
where $\int_{\mathcal{S}(R)}$ denotes integration over the region $|\bm{r}_i| < R$ for $i=1,\ldots,N$.

Taken as an isolated system, the appropriate description is the microcanonical ensemble,
\begin{equation}
P(\bm{R}, \bm{P}|E) = \frac{\delta(E-K(\bm{P})-\Phi(\bm{R}))}{\Omega(E)}
\end{equation}

where the configurational distribution is obtained by integration over the momenta,
\begin{align}
P(\bm{R}|E) & = \int d\bm{P} P(\bm{R}, \bm{P}|E) \nonumber \\
            & = \int d\bm{P} \left[\frac{\delta(E-K(\bm{P})-\Phi(\bm{R}))}{\Omega(E)}\right] \nonumber \\
            & = \int dK \Omega_K(K)\frac{\delta(E-K-\Phi(\bm{R}))}{\Omega(E)} \nonumber \\
            & = \frac{\Omega_K(E-\Phi(\bm R))}{\Omega(E)}.
\end{align}

Replacing the definition of $\Omega_K$ in Eq. \ref{eq_omega_k} we have
\begin{equation}
P(\bm{R}|E) = \frac{1}{\eta(E)}\left[E-\Phi(\bm{R})\right]_+^{\frac{3N}{2}-1},
\label{eq_micro_coord}
\end{equation}
where $[x]_+ = x$ for $x >= 0$, zero otherwise. The normalization constant $\eta(E)$ is given by
\begin{equation}
\eta(E) = \int d\phi\;\mathcal{D}(\phi) \;\left[E-\phi\right]_+^{\frac{3N}{2}-1}.
\label{eq_eta}
\end{equation}

Using the CDOS it is possible to write the microcanonical probability density of $\phi$ as
\begin{equation}
P(\phi|E) = \frac{1}{\eta(E)}\left[E - \phi\right]_+^{\frac{3N}{2}-1} \mathcal{D}(\phi).
\label{eq_prob_phi}
\end{equation}

The entropy $S(E)$ can also be expressed~\cite{Davis2011} in terms of $\eta(E)$ in Eq. \ref{eq_eta},
\begin{equation}
S(E) = S_0 + k_B\ln \eta(E).
\end{equation}

By differentiation with respect to $E$, it follows that
\begin{equation}
\frac{\partial}{\partial E}\ln \eta(E) = \beta(E) = \Big<\frac{3N-2}{2(E-\phi)}\Big>_E,
\label{eq_beta}
\end{equation}
which gives us the kinetic estimator of the inverse temperature,
\begin{equation}
\beta_K(\phi) \defeq \frac{3N-2}{2(E-\phi)},
\label{eq_beta_kin}
\end{equation}
such that $\big<\beta_K\big>_E=\beta(E)$. In a similar way as the microcanonical inverse temperature $\beta(E)$ is defined in terms of the full density of states $\Omega(E)$,
it is possible to define an inverse temperature from the CDOS, namely the \emph{configurational inverse temperature}, as
\begin{equation}
\beta_\mathcal{D}(\phi) \defeq \frac{\partial}{\partial \phi}\ln \mathcal{D}(\phi),
\label{eq_conf_beta}
\end{equation}
for which it holds that $\big<\beta_\mathcal{D}\big>_E = \big<\beta_K\big>_E = \beta(E)$. This can be seen from the conjugate variables theorem~\cite{Davis2012,Davis2016c} applied to the
potential energy distribution in Eq. \ref{eq_prob_phi},
\begin{align}
\Big<\frac{\partial \omega}{\partial \phi}\Big>_E & = -\Big<\omega(\phi)\frac{\partial}{\partial \phi}\ln P(\phi|E)\Big>_E \nonumber \\
    & = \Big<\omega(\phi)\Big(\beta_K(\phi)-\beta_\mathcal{D}(\phi)\Big)\Big>_E,
\end{align}
for $\omega=\omega(\phi)$ an arbitrary, differentiable function of $\phi$. For the choice $\omega(\phi)=1$, we have
\begin{equation}
\Big<\beta_K\Big>_E = \Big<\beta_\mathcal{D}\Big>_E.
\end{equation}

\section{Computational methods}
\label{sect_comput}

For the calculation of the density of states, in this case $\mathcal{D}(\phi)$, several methods exist. One of the most widely known is the Wang-Landau
procedure~\cite{Wang2001}, in which a random walk is performed in configuration space through the Metropolis algorithm, with
acceptance probability given by
\begin{displaymath}
p_\text{acc}(\bm{R} \rightarrow \bm{R}') = \min\left(1, \;\frac{\mathcal{D}(\Phi(\bm{R}))}{\mathcal{D}(\Phi(\bm{R}'))}\right).
\end{displaymath}

This achieves a flat distribution of energies as the Markov Chain dynamics converges to the generalized ensemble $$\rho(\phi) \propto 1/\mathcal{D}(\phi).$$ Because the
value of $\mathcal{D}(\phi)$ is not known a priori, the procedure starts with an initial guess (usually uniform) and updates it for every visited energy $\phi_i$ using the rule
$$\mathcal{D}(\phi_i) \rightarrow \mathcal{D}(\phi_i)\cdot f,$$ where $f$ is a factor which is decreased according to some predefined schedule, usually by the
rule $f_{i+1} = \sqrt{f_i}$ so it converges to 1.

Microcanonical simulations in the ensemble defined by Eq. \ref{eq_micro_coord} can be performed via Monte Carlo Metropolis as proposed by J. R. Ray~\cite{Ray1991}, in which the
acceptance probability becomes
\begin{equation}
p_\text{acc}(\bm{R} \rightarrow \bm{R}') = \min\left(1, \;\left[\frac{E-\Phi(\bm{R'})}{E-\Phi(\bm{R})}\right]^{\frac{3N}{2}-1}\right),
\label{eq_pacc_micro}
\end{equation}
instead of the usual
\begin{equation}
p_\text{acc}(\bm{R} \rightarrow \bm{R}') = \min\left(1, \exp(-\beta \Delta\phi)\right),
\label{eq_pacc_canon}
\end{equation}
employed in canonical Metropolis simulation. Note that, for $|\Delta \phi| \ll E-\Phi$ where $\Delta \phi = \Phi(\bm{R}')-\Phi(\bm{R})$, we can provide the following
convenient approximation,
\begin{align}
\left[\frac{E-\Phi(\bm{R}')}{E-\Phi(\bm{R})}\right]^{\frac{3N}{2}-1} & = \left[\frac{E-\Phi(\bm{R})-\Delta \phi}{E-\Phi(\bm{R})}\right]^{\frac{3N}{2}-1} \nonumber \\
                                         & = \left[1-\frac{\Delta \phi}{E-\Phi(\bm{R})}\right]^{\frac{3N}{2}-1} \nonumber \\
                                         & = \exp\Big(\frac{3N-2}{2}\ln \Big(1-\frac{\Delta \phi}{E-\Phi(\bm{R})}\Big)\Big) \nonumber \\
                                         & \approx \exp(-\beta_K(\Phi(\bm{R}))\Delta \phi).
\end{align}

This means that, for small proposed displacements, microcanonical Metropolis sampling can be treated as a canonical Metropolis sampling with variable inverse temperature, given
by $\beta_K$ in Eq. \ref{eq_beta_kin}. In the case of vanishing potential energy fluctuations, the microcanonical ensemble predictions coincide with the canonical predictions
at $\beta=\beta(E)$.

\begin{figure}[b!]
\begin{center}
\includegraphics[scale=0.56]{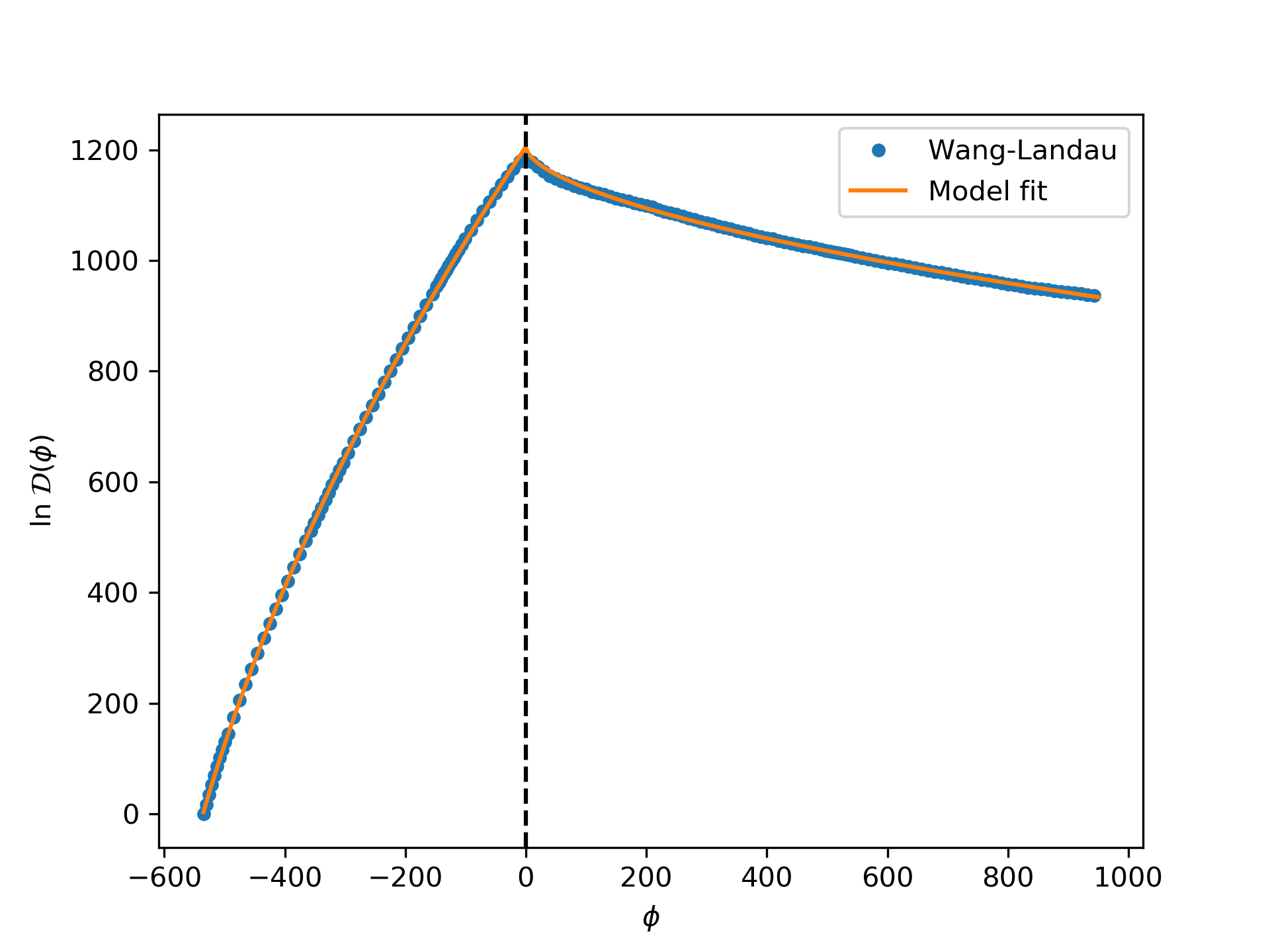}
\end{center}
\caption{Logarithm of the configurational density of states $\mathcal{D}(\phi)$ for a Coulomb system confined to a sphere, for energies $\Phi$ between
-535 $\phi_0$ and 950 $\phi_0$, as computed by the Wang-Landau algorithm. The dashed line indicates $\phi$=0, value above which the curvature changes.}
\label{fig_cdos}
\end{figure}

\section{Results}
\label{sect_results}

For the system of $N$ particles interacting via $\Phi$ in Eq. \ref{eq_phi_spin} inside a sphere of radius $R$, the logarithm of the CDOS calculated using the Wang-Landau algorithm is shown
in Fig. \ref{fig_cdos}. It can be seen that the CDOS is asymmetrical, having an inflection point exactly at $\phi$=0 which divides two regions with different curvature. These regions
can be described by the simple empirical model,
\begin{equation}
\ln \mathcal{D}(\phi) = \begin{cases}
a + A(\phi+b)^\alpha, \quad \text{if}\; \phi < 0,\\
a + A\cdot b^\alpha - B\phi^\alpha, \quad \text{if}\; \phi \geq 0, \\
\end{cases}
\end{equation}
with parameters $a$=-596.0357, $b$=626.8341, $A$=44.9553, $B$=5.3482 and $\alpha$=0.5731.

\begin{figure}
\begin{center}
\includegraphics[scale=0.56]{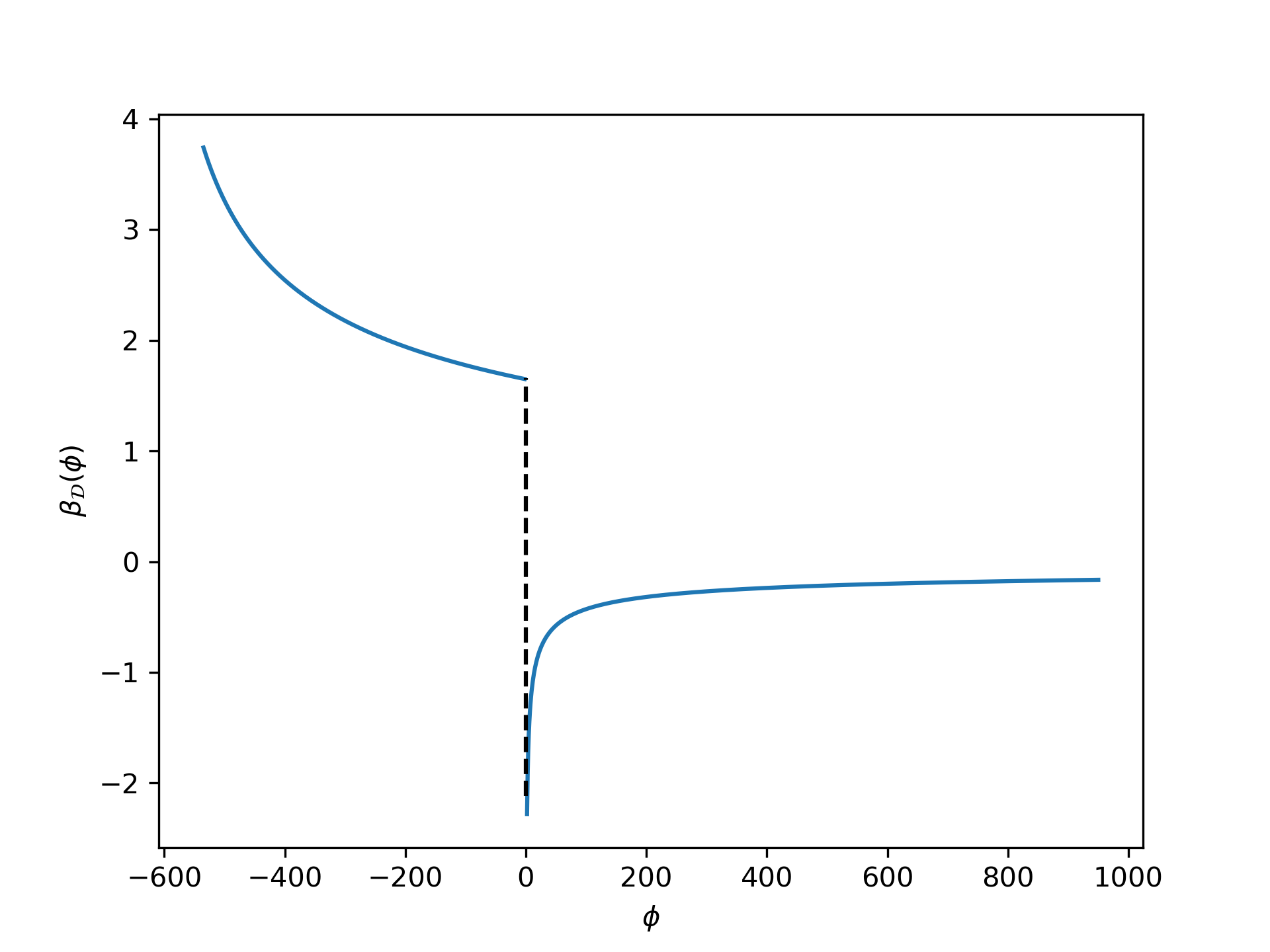}
\end{center}
\caption{Configurational inverse temperature $\beta_\mathcal{D}(\phi)$ obtained from the CDOS in Fig. \ref{fig_cdos}.}
\label{fig_beta_D}
\end{figure}

The configurational inverse temperature $\beta_D$, defined in Eq. \ref{eq_conf_beta}, is shown in Fig. \ref{fig_cdos}. A discontinuity at $\phi=0$ can be seen, reflecting the change
in curvature of the CDOS. Because $\beta_D < 0$ for $\phi > 0$, no macroscopic state in the microcanonical ensemble is compatible with strictly positive potential energies, as this
would imply $$\big<\beta_D\big>_E = \big<\beta_K\big>_E < 0,$$ which is incompatible with the definition of $$\beta_K(\phi) = \frac{3N}{2(E-\phi)} > 0.$$ Consequence of this is the
fact that the potential energies tend to ``pile up'' towards $\phi=0$ for high enough total energies, as can be seen in the lower panel of Fig. \ref{fig_micro} for energies up to
$E$=3000 $\phi_0$. In other words, despite the fact that microscopic states with $\phi$ above 900 $\phi_0$ do exist, as shown by the CDOS in Fig. \ref{fig_cdos}, they are not
accessible in microcanonical conditions.

\subsection{Thermodynamical properties}

\begin{figure}
\begin{center}
\includegraphics[scale=0.56]{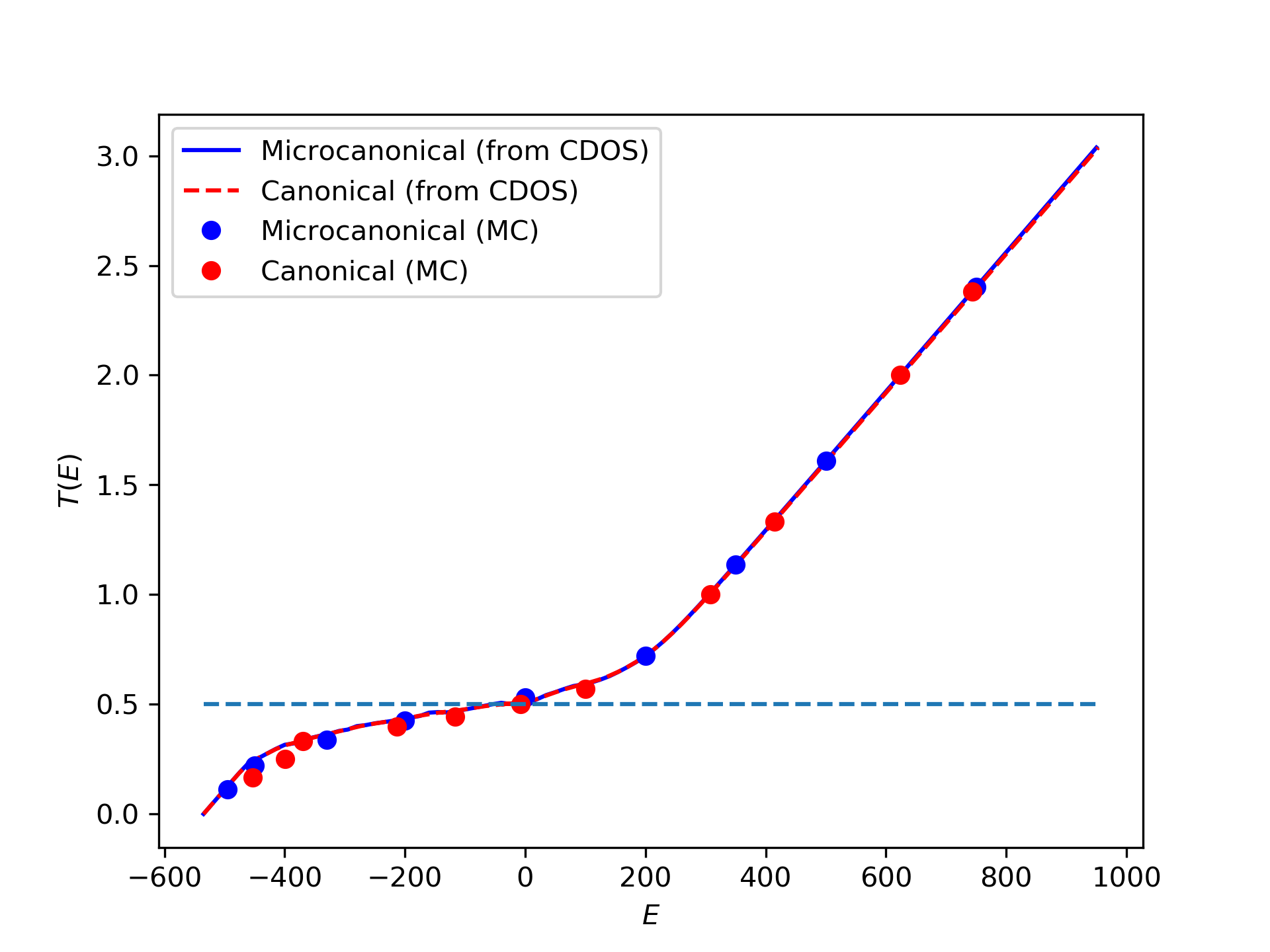}
\end{center}
\caption{Caloric curve $T(E)$ for a Coulomb system confined to a sphere, for energies $E$ between -535 $\phi_0$ and 950 $\phi_0$ units. Dots represent microcanonical (blue)
and canonical (red) Metropolis simulations while the solid and dashed lines are microcanonical and canonical predictions, respectively, based on the CDOS in Fig. \ref{fig_cdos}.
The horizontal (dashed) line corresponds to the temperature $T$=0.5 $T_0$ at $E$=0.}
\label{fig_caloric}
\end{figure}

Using the microcanonical Monte Carlo method defined by the acceptance probability in Eq. \ref{eq_pacc_micro}, we have computed the caloric curve $T(E)$ by collecting the average
$$\frac{1}{k_B T(E)} = \Big<\frac{3N-2}{2(E-\Phi)}\Big>_E,$$ and compared with the predictions of Eqs. \ref{eq_beta} and \ref{eq_prob_phi} with the CDOS in Fig. \ref{fig_cdos}.
This caloric curve is shown in Fig. \ref{fig_caloric}, together with independent canonical Monte Carlo simulations, using the acceptance probability in Eq. \ref{eq_pacc_canon} instead.

\begin{figure}
\begin{center}
\includegraphics[scale=0.56]{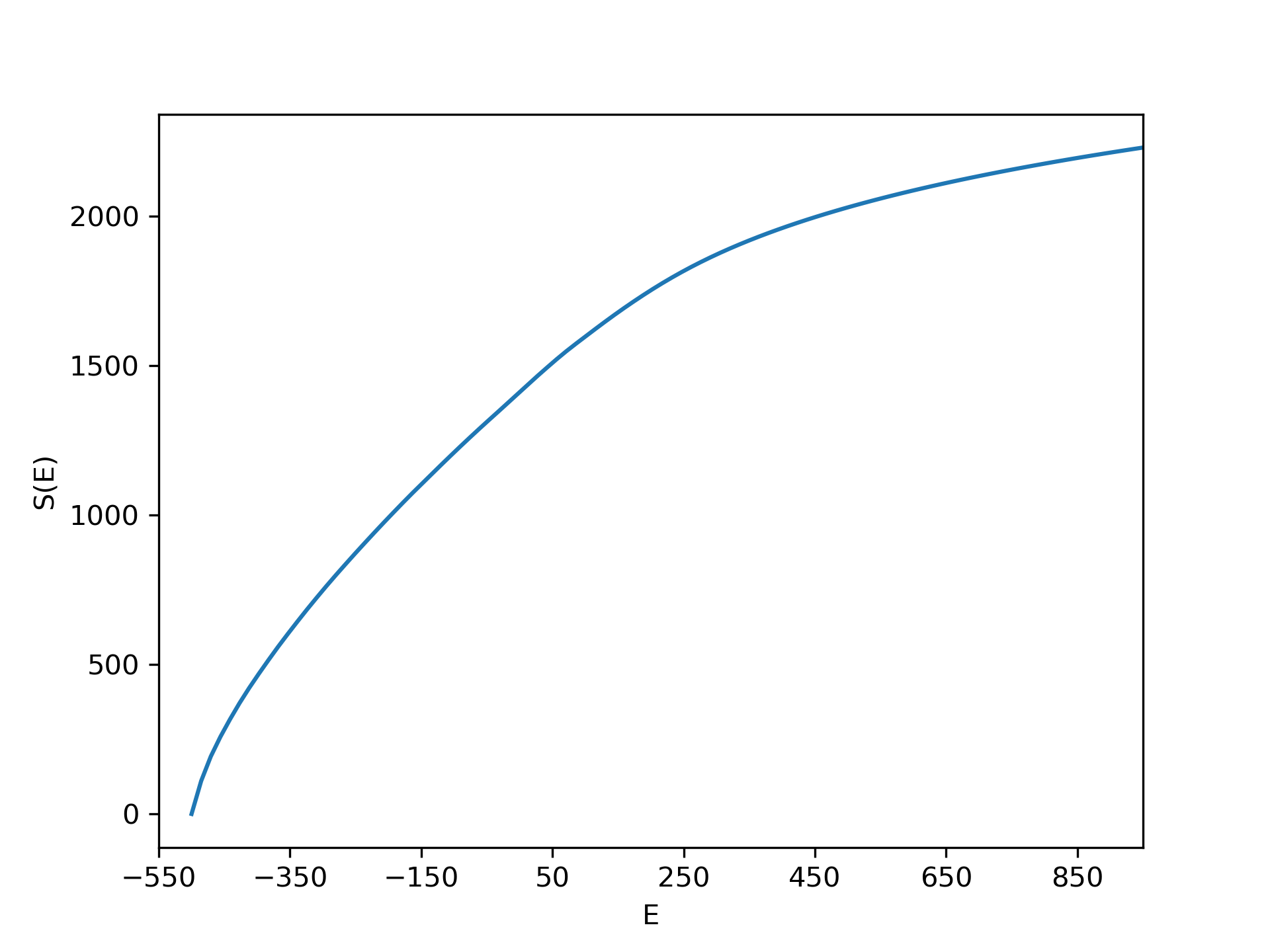}
\end{center}
\caption{Total entropy for a Coulomb system confined to a sphere, for energies $E$ between -535 $\phi_0$ and 950 $\phi_0$.}
\label{fig_entropy}
\end{figure}

Complete equivalence between both ensembles is seen, despite the fact that a transition seems to occur between two branches with nearly constant specific heat, around $T^*$=0.5 $T_0$, the
temperature corresponding to $E$=0. The continuous behavior of the caloric curve is verified in the microcanonical entropy $S(E) = k_B\ln \eta(E)$, shown in Fig.
\ref{fig_entropy}. The entropy is monotonically increasing with $E$ without any ``backbending'', thus despite the discontinuity of the configurational inverse temperature there is no
evidence in this system of a first-order phase transition, as commonly occurs in long-range interacting~\cite{Campa2009} and small systems~\cite{Eryurek2007}.

\newpage

\begin{figure}[t!]
\begin{center}
\includegraphics[scale=0.56]{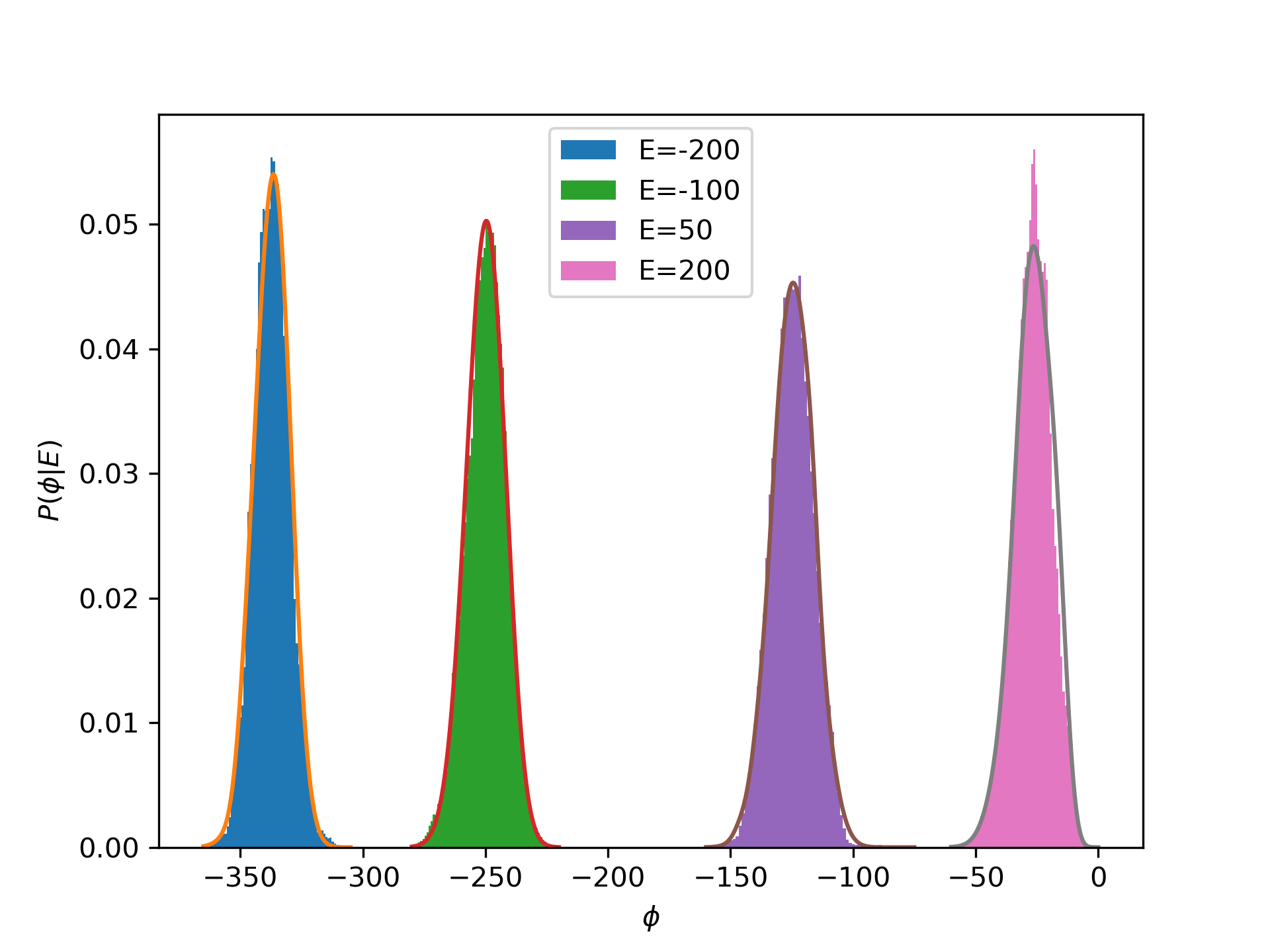}
\includegraphics[scale=0.56]{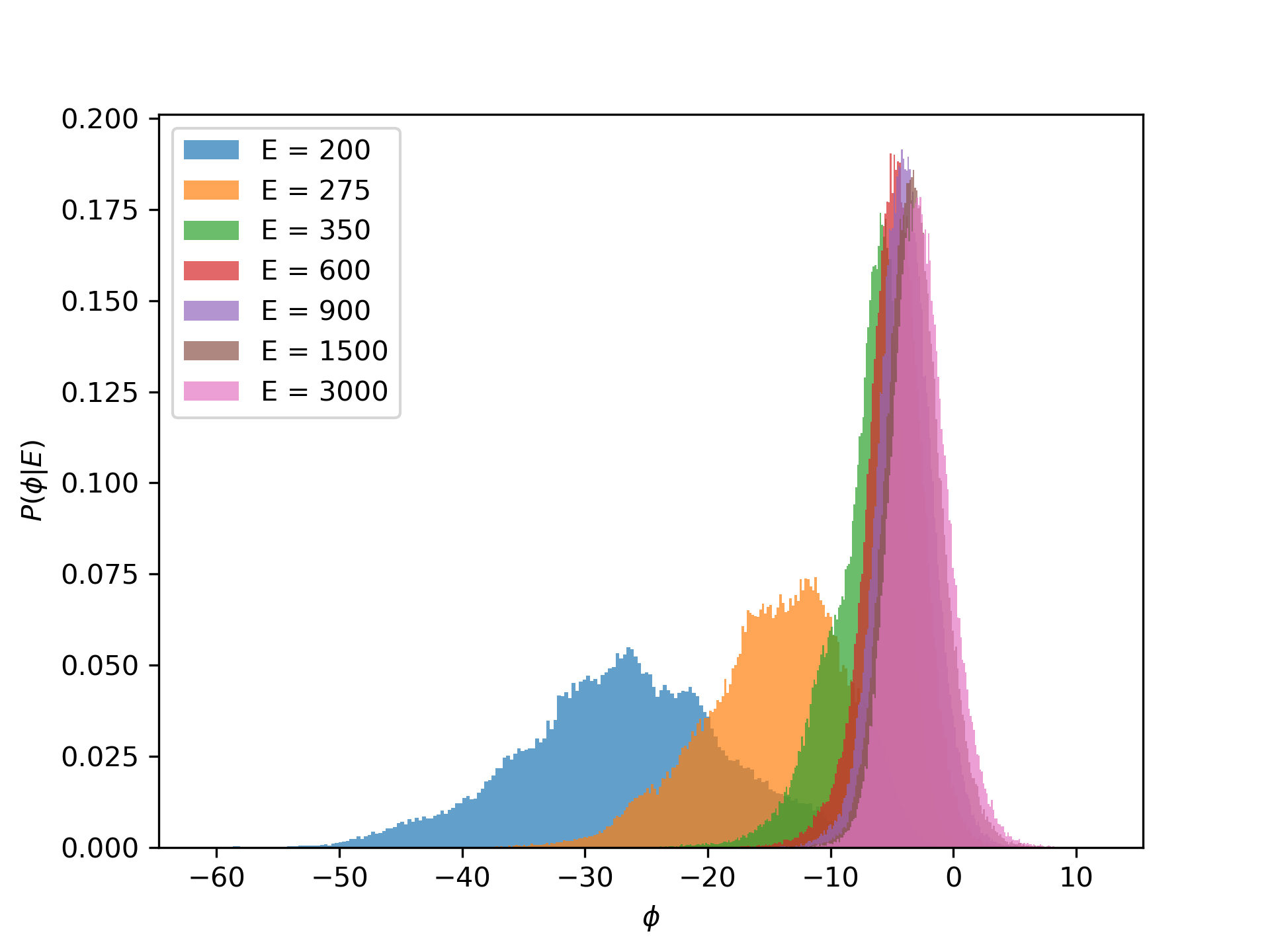}
\end{center}
\caption{Upper panel, potential energy histograms from microcanonical Monte Carlo simulation, from $E$=-200 $\phi_0$ to $E$=200 $\phi_0$. The solid lines are the corresponding
probability densities predicted using Eq. \ref{eq_prob_phi}. Lower panel, potential energy histograms for microcanonical simulation for total energies between $E$=200 $\phi_0$
and $E$=3000 $\phi_0$, showing the concentration of the distribution mass towards $\phi=0$.}
\label{fig_micro}
\end{figure}

\begin{figure}[h!]
\begin{center}
\includegraphics[scale=0.58]{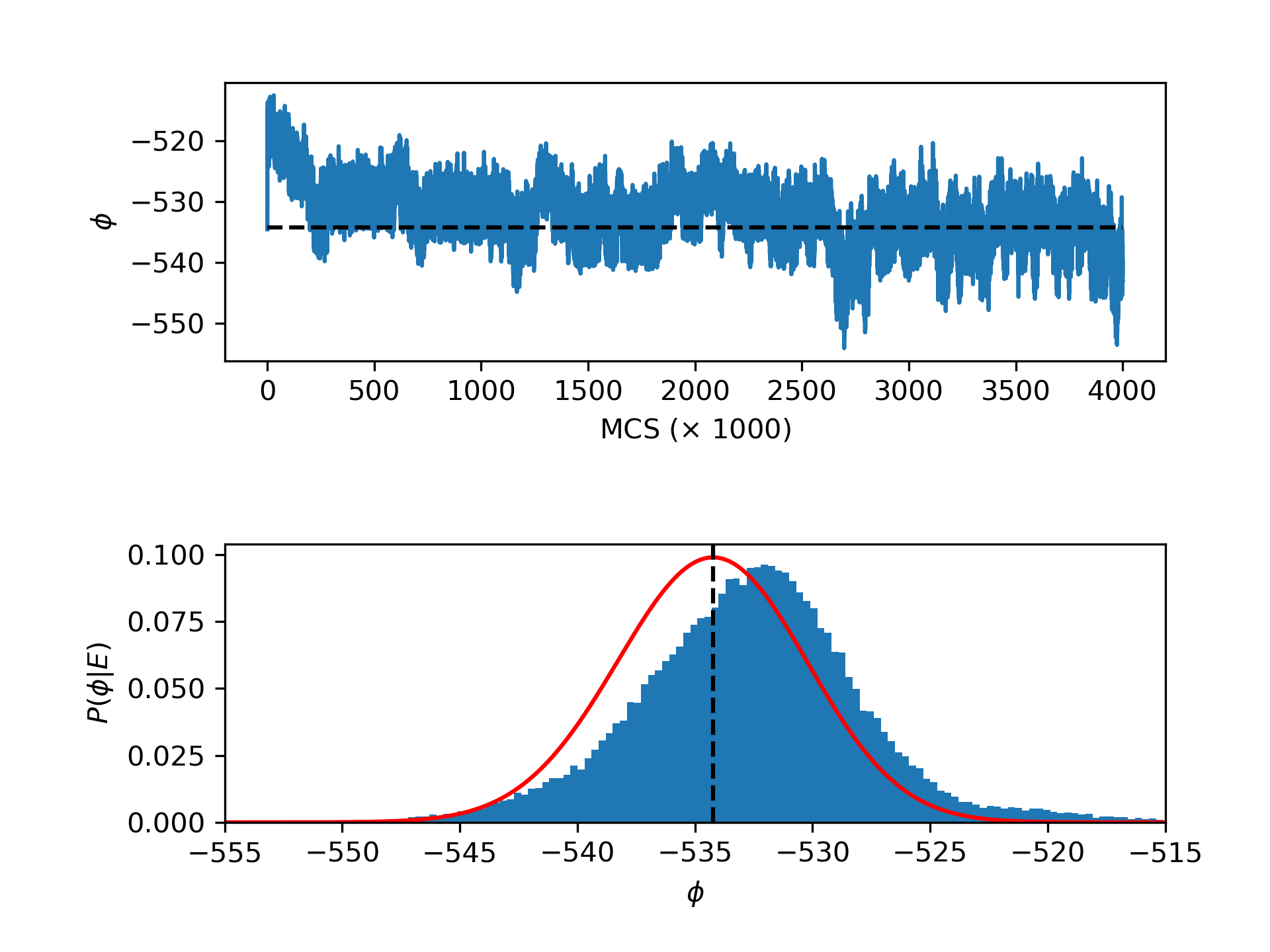}
\end{center}
\caption{Upper panel, potential energy as a function of Monte Carlo steps in the microcanonical ensemble, for $E$=-450 $\phi_0$. Lower panel, histogram of the potential energy collected
in the same simulation. Due to the dynamic multistability phenomenon observed, the convergence of the histogram to the correct distribution (red solid line) given by Eq.
\ref{eq_prob_phi} is much slower than for other energies.}
\label{fig_traces}
\end{figure}

In order to assess the accuracy of the CDOS computed using the Wang-Landau algorithm beyond averages, we computed the microcanonical potential energy distributions according
to Eq. \ref{eq_prob_phi} for several total energies, and compared them with empirical histograms collected from microcanonical Monte Carlo simulation. These results are shown in
Fig. \ref{fig_micro}. We can see that, in all cases, the CDOS is capable of describing correctly the shape of the potential energy fluctuations, at least above $E$=-200 $\phi_0$.
In the lower branch of the caloric curve, however, an interesting phenomenon of dynamical multistability is observed in the microcanonical Monte Carlo simulation, as shown for
$E$=-450 $\phi_0$ in Fig. \ref{fig_traces}. In this case, even when the simulation starts from the most probable potential energy, given by the condition
$$\frac{\partial}{\partial \phi}\ln P(\phi|E)\Big|_{\phi=\phi^*}=0,$$ that is,
\begin{equation}
\beta_K(\phi^*) = \beta_\mathcal{D}(\phi^*),
\end{equation}
which has solution $\phi^*$=-534.231 $\phi_0$ for $E$=-450 $\phi_0$ (shown as the black dashed line in Fig. \ref{fig_traces}) the system transits between several macroscopic
states, with considerable lifetimes. This effect resembles the behavior of glasses~\cite{Ghosh2006} and proteins~\cite{Wust2011}, which commonly have a complex potential energy
landscape that makes direct sampling difficult. The dynamical multistability phenomenon reduces the efficiency of microcanonical sampling for low energies, as we have to wait much
longer times for the system to explore the different macroscopic states with the correct frequency, thus requiring extremely long simulations to collect reliable statistics.

\section{Concluding Remarks}
\label{sect_conclud}

We have computed thermodynamic properties of a system of unscreened charged particles confined into a spherical region by using one of the well-established flat-histogram methods,
the Wang-Landau algorithm. Our system, while conceptually simple, is still challenging from the point of view of computational statistical mechanics, as we have shown in direct microcanonical Metropolis sampling. We have determined the thermodynamics of the system, in particular we report the configurational density of states, the full thermodynamic
entropy, the caloric curve and the microcanonical potential energy distributions.

The shape of the configurational inverse temperature easily explains a concentration of the potential energy distribution mass around $\Phi = 0$ at high total energies, which makes
most states with $\Phi > 0$ inaccessible from the microcanonical ensemble. On the other hand, the presence of a dynamical multistability phenomenon complicates the direct
microcanonical sampling at low energies, highlighting the usefulness of the generalized-ensemble approaches to computational statistical mechanics.

\section*{Acknowledgements}
The authors gratefully acknowledge funding from Anillo ACT-172101 grant.


\end{document}